\documentclass{amsart}
\usepackage{a4wide}
\usepackage[dvips]{graphicx}
\usepackage{bbm}
\usepackage{subfigure}
\usepackage{cite}

\def\E{\mathbb{E}}

\def\P{\mathbb{P}}

\def\Scal{\mathcal{S}}

\def\cal{\mathcal}

\newcommand{\ind}[1]{\mathbbm{1}_{\{#1\}}}
\newtheorem{theorem}{Theorem}

\newtheorem{defi}{Definition}
\newtheorem{lemma}{Lemma}
\newtheorem{proposition}{Proposition}

\usepackage{color}

\long\def\symbolfootnote[#1]#2{\begingroup%
\def\thefootnote{\fnsymbol{footnote}}\footnote[#1]{#2}\endgroup}

\usepackage{fancyhdr}
\title[Channel Fragmentation in Dynamic Spectrum Access Systems]{Channel Fragmentation in Dynamic Spectrum Access Systems - a Theoretical Study}

\author[E. Coffman]{Ed Coffman}
\address[E. Coffman,S. Tarami, G. Zussman]{Columbia University, Department of Electrical Engineering, New York, NY 10027, USA}
\email{egc@ee.columbia.edu}

\author[Ph. Robert]{Philippe  Robert}
\address[Ph. Robert,F. Simatos]{INRIA Paris --- Rocquencourt,  Domaine de Voluceau, 78153 Le Chesnay, France.}
\email{Philippe.Robert@inria.fr}
\urladdr{http://www-rocq.inria.fr/\string~robert}

\author[F. Simatos]{Florian Simatos}
\email{Florian.Simatos@inria.fr}

\author[S. Tarumi]{Shuzo Tarumi}
\email{shuzo@ee.columbia.edu}

\author[G. Zussman]{Gil Zussman}
\email{gil@ee.columbia.edu}

\begin{document}

\begin{abstract}
Dynamic Spectrum Access systems exploit temporarily available
spectrum (`white spaces') and can spread transmissions over
a number of non-contiguous sub-channels. Such methods are highly beneficial in terms of spectrum utilization.
However, excessive fragmentation degrades performance and hence
off-sets the benefits. Thus, there is a need 
to study these processes so as to determine how
to ensure acceptable levels of fragmentation. Hence, 
we present experimental and analytical results derived from
a mathematical model. We model a system operating at
capacity serving requests for bandwidth by assigning a
collection of gaps (sub-channels) with no limitations on the fragment
size. Our main theoretical result shows
that even if fragments can be arbitrarily small, the system does
not degrade with time. Namely, the average total number of fragments
remains bounded.  Within the very difficult class of dynamic
fragmentation models (including models of
storage fragmentation), this result appears to be the first
of its kind. Extensive experimental results describe
behavior, at times unexpected, of fragmentation under 
different algorithms.
Our model also applies to dynamic linked-list
storage allocation, and provides a novel analysis in that domain. We
prove that, interestingly, the 50\% rule of the classical
(non-fragmented) allocation model carries over to our  model.
Overall, the paper provides insights into the potential
behavior of practical fragmentation algorithms.
\end{abstract}

\maketitle

\section{Introduction}\label{sec:intro}
This paper  focuses on dynamic resource allocation algorithms in Dynamic Spectrum Access  Networks (also known as Cognitive Radio Networks). A Cognitive Radio is a concept that was
first defined by Mitola \cite{Mitola2001,Mitola2000} as a radio that
can adapt its transmitter parameters to the environment in which it
operates. Technically, it is based on the concept of Software Defined Radio
\cite{BoseSI1998} that can alter parameters such as frequency band,
transmission power, and modulation scheme through changes in
software. According to the Federal Communications Commission (FCC), a large
portion  of the assigned spectrum   is
used only sporadically \cite{FCC03222}.  Because  of their adaptability
and capability to utilize the  wireless spectrum opportunistically, algorithms for Dynamic
Spectrum Access are key enablers to efficient use of the  spectrum. 
Hence, the potential of Cognitive Radio Networks has  been recently identified  by various
policy \cite{FCC0557,FCC08260},  research \cite{DARPAF2003,DARPAV2003}, and
standardization \cite{IEEE80222,IEEE41,ETSI} organizations.

\begin{figure}[t]
\center
\scalebox{0.5}{\includegraphics{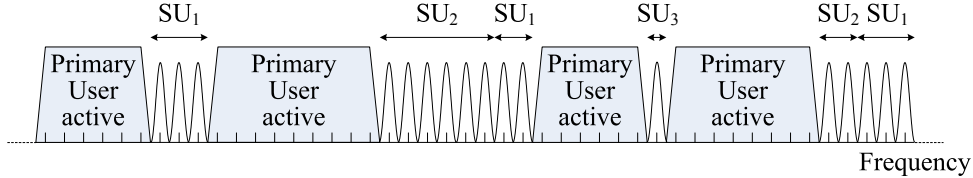}}
\caption{Non-contiguous OFDM with Primary and Secondary Users (SUs), where the Secondary Users use non-contiguous channels that do not overlap with the Primary Users' channels.}
\label{fig:OFDMA}
\end{figure}

Under the basic model of  Cognitive Radio Networks \cite{AkyildizLVM2006}, Secondary Users (SUs)
can  use \emph{white  spaces} (also known as \emph{spectrum holes}) that are  not  used by  the Primary  Users  but must  avoid
interfering  with active Primary  Users (e.g.,  Figure~\ref{fig:OFDMA}). For  example, the
Primary and Secondary Users can be  viewed as TV broadcasters and cellular operators using
available  TV bands \cite{IEEE80222}.
Under this  model, one assumes  that when a  transmission of a  Primary User takes  place, it
occupies a  predefined band.  An SU may identify  spectrum holes (not  used by
Primary or other  Secondary Users) and can  allocate its bandwidth  among a  number of subchannels,
occupying a number of holes (not necessarily contiguous).  This can  be realized, for example, by employing a variant of
Orthogonal  Frequency-Division  Multiplexing  (OFDM)   that  is  capable  of  deactivating
sub-carriers   which  have   the  potential   to  cause   interference  to   other  users
\cite{GhasemiS2008,MahmoudYA2009,Horne2005,JondralW2004,QinetiQ2006,Rajbanshi2007} (such a non-contiguous OFDM scheme is shown in 
Figure \ref{fig:OFDMA}).

The use of non-contiguous  bandwidth blocks results in non-trivial behavior  even for very simple
scenarios. As an example, due to the dynamic use of the available spectrum holes
and the arrivals and departures of SU bandwidth requests, SUs may need to transmit
in a set of smaller and smaller holes. This will lead to a highly fragmented spectrum whose maintenance may require complex algorithmic solutions.
Although the practical (physical layer) aspects of OFDM-based Dynamic Spectrum
Access have been extensively studied  recently, the use of
fragmented (i.e., non-contiguous)  spectrum  introduces   several  new   problems
\cite{JiaZS2008,YuanBahl2007,QinetiQ2006} that significantly  differ from classical Medium
Access Control (MAC) and fragmentation problems. 

In this  paper, we study  the most \emph{basic theoretical model} in which the  spectrum is shared  by SUs only.
Those users have to transmit and receive data, and accordingly need some bandwidth for given amounts of time. Hence, bandwidth requests of SUs are characterized by a  desired total bandwidth
and the  duration of a  time interval over  which it is needed. The data transmission can take place
over a non-contiguous channel (i.e.,  a number of subchannels).  Once a transmission   terminates,  some  fragments   (subchannels)  are  vacated,  and  therefore,  gaps
(spectrum  holes/white spaces)  develop randomly in  both size and  position.

\pagestyle{plain}

\begin{figure*}[!t]
 \centering
 \includegraphics[width=5in]{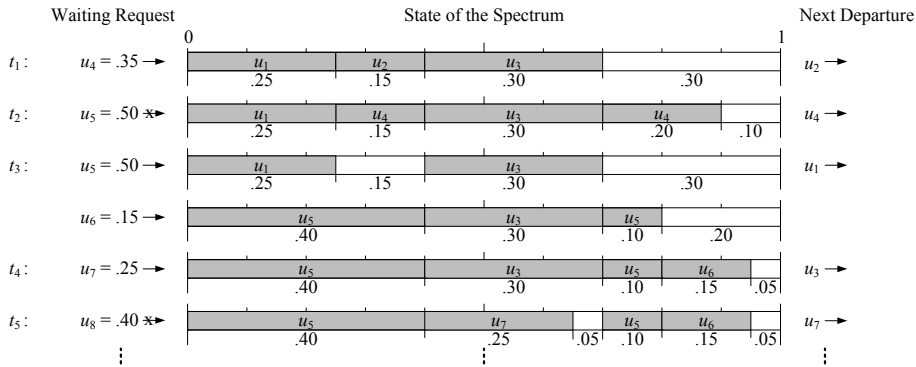}
\vspace*{-0.3cm}
 \caption{An example of the admission and departure processes and the resulting spectrum fragmentation.}
 \label{Example1}
\end{figure*}

When
allocating a channel to a new  SU request, it is  being fragmented (in the  frequency domain)
into available gaps  until the full requested  bandwidth is provided.  This  process repeats itself,
until the  next request fails  to fit into  the available fragments.  Figure~\ref{Example1}
demonstrates the  process of a transmission termination followed by the  fragmentation of waiting bandwidth requests (more details about this example can be found in
Section \ref{sec:model}). We note that from a practical point of view, such a system can be viewed as a simplistic version of an OFDM-based access-point/base-station/spectrum-broker that tracks the available spectrum holes and allocates non-contiguous bandwidth blocks to SUs.

The main  goal of the paper  is to investigate  the phenomena of fragmentation  induced by
spectrum  allocation  algorithms.  For  this   purpose,  we  will ignore  the particularities of techniques such as
OFDM and make a couple of assumptions (for a complete system description,
see Section \ref{sec:model}): (i) the system operates at capacity and there is always a waiting bandwidth request; 
and (ii) the
fragment size  is not bounded from  below. Making the first assumption allows us to study the effect of fragmentation in the worst case. Clearly, if there  are idle periods,  when there are no waiting requests,  only departures
occur and the fragmentation level of the system decreases during these periods. Similarly, the latter assumption allows us to study the system performance when artificial lower bounds on the fragment size are not imposed. This differs from OFDM-based systems in which a subcarrier  has a given  minimal bandwidth.


\subsection*{Main Results}
At first glance, it seems that given assumption (ii) above, as time passes, the fragments used by bandwidth requests might well become progressively narrower and that the number of fragments grows without bound. Clearly, such an operation model cannot be supported
by a realistic system (e.g., an OFDM-based system). However, our
analytical and experimental results show that \emph{this is not the case}.

In particular, under our model, a portion  of the  spectrum (represented by the interval  $[0,1]$) is
available  for  allocation to  requests whose  sizes 
are uniform on  $(0,\alpha]$ ($0 < \alpha \leq 1$).  Once  allocated, requests are satisfied (leave the system) after an  exponential time. A stochastic model  of channel fragmentation
is investigated which includes the asymptotic
behavior of the number  $G(t)$ of \emph{gaps}, $F(t)$ of  \emph{fragments}, and $R(t)$ of requests being served (also referred to as allocated \emph{channels}). A
Markovian description of  this system is clearly more complicated  than the vector $(G(t),
F(t),  R(t))$, since for  each of  the $R(t)$  channels, the  location and  size of  each  of its
fragments should be included in such a description.

Let $(t_n)$ be the  sequence of  request departure times. Our main result shows that the total  number of gaps and fragments
at these epochs has bounded exponential moments,
\[
\sup_{n\geq 1}\E\left(e^{\eta(F(t_n)+G(t_n))}\right)<+\infty,
\]
for some $\eta>0$. 
It follows that the sum $F(t_n)+G(t_n)$ is strongly concentrated near the origin,
indicating that, \emph{with high probability, low fragmentation of the spectrum holds under high traffic intensity}.
In our  analysis, several basic ingredients  are used: relations between  the number of
fragments and gaps in the spectrum; a drift relation of Lyapunov type for
the total number of gaps, fragments, and requests (Proposition~\ref{prop:lyapunov});
a general inequality for Markov chains (Theorem~\ref{thm:hajek});
and a  stability result of \cite{KipnisR1990}.

In addition to the analytical results, and in order to gain insight into the performance of the system, we present the results of an extensive simulation study. We show that although for a given maximum request size ($\alpha$), the number of fragments has a finite expected value, there is a linear relationship between 1/$\alpha$ and the expected number of fragments into which a request is divided. This indicates that when the requests are very small, they are also fragmented into a relatively large number of fragments. From a practical point of view, this implies that \emph{imposing a lower bound on the fragment size is important} in order to reduce the complexity resulting from maintaining a channel heavily fragmented.

We show that different bandwidth allocation algorithms have significantly different performance in terms of the fragmentation occurring in the system. In particular, an algorithm (referred to as LFS) that allocates gaps in decreasing order of their sizes reduces the fragmentation by almost an order of magnitude compared to other algorithms. Interestingly, we also
show that the number of fragments is distributed according to a
Normal law with a relatively low mean value. Finally, we observe an
unexpected reappearance of the {\em 50\% rule} \cite{Knuth1997},
according to which, on average the number of gaps (unoccupied
fragments) is half the number of requests being served (this
rule differs from the original 50\% rule by Knuth
\cite{Knuth1997} which is briefly discussed below).

\subsection*{Related Work}\label{sec:related}

The areas of Dynamic Spectrum Access and Cognitive Radio Networks have been extensively studied (for a comprehensive review of previous work see
\cite{AkyildizLVM2006,AkyildizLC2009,KrenikWD2007,MahonenZ2007}). In particular, problems stemming from fragmented  spectrum  have been considered in 
\cite{JiaZS2008,YuanBahl2007,QinetiQ2006}. In addition, several previous works focused on the technical aspects of channel fragmentation \cite{GeirhoferTS2008,GhasemiS2008,Horne2005,HouYZSZ2009,JondralW2004,MahmoudYA2009,Rajbanshi2007}. To the best of our knowledge, despite these recent efforts, the fragmentation problem considered in this paper has not been studied before.


Yet, the  spectrum allocation  problems described here  can  be
characterized in  the context of classical dynamic  storage allocation
problems of computers,  see Knuth~\cite{Knuth1997}. In that context,
the term ``spectrum" refers to the  storage unit, ``bandwidth"
refers to storage, and ``channel" refers  to a region in storage
containing a file.  In the original storage model, fragmentation
refers only to the gaps of unoccupied storage interspersed with
intervals of occupied storage: files are not fragmented.  The 50\%
rule was derived for this model and asserted that the number of gaps
was approximately half the number of files when exact fits of files
into gaps were rare.  Our notion of fragmentation applies also to
the files themselves, so our discovery that the 50\% rule continued
to apply was a surprising one at first. Note that, in terms of the
storage application, our model corresponds well to {\it linked-list}
allocation of files -- channels allocated to requests are sets of linked, disjoint
segments of storage allocated to specific file fragments. Our results provide
a novel analysis for such systems, and have
implications for the {\em garbage-collection/defragmentation}
process in linked-list systems.

We note that studies of dynamic storage allocation have been around for some
40 years, and widely recognized as posing very challenging problems
to both combinatorial and stochastic modeling and analysis. In particular, results of  the type
found in this paper, rigorous within stochastic models, seem to be quite new.

For a system without fragmentation, the early results of Kipnis and Robert \cite{KipnisR1990}
concern a stochastic analysis of the number of requests being served in the system.  Fragmentation is not an
issue in their case since request  allocations can  be moved  when  needed in  order to  put
available space all together in one  block.   A major result in  \cite{KipnisR1990} applying to our
model  asserts the existence  and uniqueness  of an  invariant measure  for the  number of
requests.  Explicit  formulas   for  our  system  are  hard  to  come   by,  but  those  in
\cite{KipnisR1990} for  the maximal departure rate  in special cases  have provided useful
checks on our experiments.

\subsection*{Organization of the Paper}
The remainder of this paper  is organized as follows.  Section~\ref{sec:model} describes the system, allocation algorithm, and the mathematical variables describing fragmentation. Section~\ref{sec:expr} presents simulation results that bring out the behavior of the system and the effects of fragmentation.  Section~\ref{sec:behav}
introduces some relations between the variables describing the fragmentation and a  {\em 50\% limit law} for  the relation between
the number of active  channels and the number of gaps in  the spectrum at departure times.
Section~\ref{sec:stab} contains our main mathematical result, Theorem~\ref{thm:sup}, which
shows  that the  average value  of the  total number  of fragments  and gaps  is bounded.
Furthermore, in some  cases, the existence of an  equilibrium distribution is established in
Theorem~\ref{thm:stab}. Section~\ref{sec:algorithms} discusses  algorithmic  issues, such  as  changes  in
performance resulting from alternative algorithms  for sequencing through the list of gaps
when  constructing  a channel  for  a  newly  admitted request.  Section~\ref{sec:normal}  presents  experiments suggesting  that
Normal approximations for the total number of fragments and gaps
hold. Section  \ref{sec:conc}
discusses the results and future research directions. 

\section{The Model}\label{sec:model}

\subsection*{System Model}
As  mentioned above, we  consider requests for bandwidth  queued up  waiting to be served, each  of them identified with the amount of  bandwidth required and a  time telling how
long it  wants a channel with that total bandwidth.
Channels are  allocated to bandwidth requests on  a FCFS (first-come-first-served)  basis, subject to
available  bandwidth; the  channels must  remain  fixed while active;  they
depart  after varying  delays,  so \emph{gaps} alternating with sequences of sub-channels (channel \emph{fragments})
develop over time.

For convenience, we  normalize the spectrum to the interval $[0,1]$,  so all bandwidth requests
are numbers in $[0,1]$. There is  a  queue of  waiting requests that never empties.  Assuming that the spectrum initially has no channels in use,    the allocation
process    begins   by   allocating   consecutive    channels   i.e.,   consecutive
subintervals of  [0,1], to  requests whose sizes are $u_1, u_2, \ldots  $ until  a request
whose size $u_i,~i>1,$ is reached which exceeds available bandwidth, i.e.,
 $$   u_1 + \cdots + u_{i-1} \leq 1 < u_1 + \cdots + u_i $$

All $i-1$ of the channels  now begin their independent delays.  Subsequent  state transitions take  place at departure  epochs when the  delays of
currently allocated channels expire. At such  epochs,
all fragments of the departing channel are released.  Suppose all requests up  to $u_j$ have been
allocated  channels and  a request  $u_i,~ i  \leq j$,  departs, releasing  its allocated
channel.   Then $u_{j+1},  u_{j+2}, \ldots  $ are  allocated their  requested bandwidths
until, once again, a  request is encountered that asks for more bandwidth than is available.
All channels then begin  or continue their delays
until the next departure. A standard rule for setting up a channel is to
scan the spectrum, from  one end to the other, with gaps  of available bandwidth allocated
to  fragments of  the  new requested  bandwidth until  enough has  been
allocated to  satisfy the entire  request. The  last gap used  in satisfying a  request is
normally only partially used; the partial  allocation in the last gap is left-justified in
that gap. We refer to this scanning rule as {\em  Linear Scan}.

An example is  shown in Figure~\ref{Example1}.  Allocations for the  first 8 user requests whose sizes are
$u_1,  \cdots ,u_8$ are  shown, assuming  that, at  time 0,  the spectrum  is not  in use.
The requests with sizes $u_1,$ $u_2$, and  $u_3$ are the first to be allocated  channels; $u_4$ must wait
for a departure, since  the first 4 request sizes sum to more than  1.  The variables $t_i$ give the
sequence of  departure times of  allocated requests. We  see that the first  occurrence of
fragmentation takes place at  the departure of $u_2$ and the subsequent  admission of $u_4$; an
initial fragment  of $u_4$  is placed in  the gap left  by $u_2$  and a final  fragment is
placed after $u_3$.   Note also that, even  after $u_2$ and $u_4$ have  departed, there is
still not enough  bandwidth for $u_5$. After the additional departure  of $u_1$, both $u_5$
and $u_6$, but not $u_7$, can be allocated bandwidth.

In  Section \ref{sec:algorithms}, we  shall also evaluate two  alternatives to the \emph{Linear Scan} rule: a \emph{Circular Scan}, by which  each linear scan starts  where the
previous  one   left  off,  and  a  \emph{Largest-First Scan}
intended  to further reduce fragmentation.  Although these algorithms make different scans of the gap sequence,  they are all alike in their treatment of the last gap occupied: {\em  the last fragment is left justified in the last gap.}  This is a key assumption, and it is very likely to hold in practice. 

The allocation process described in this section contrasts with  the classical model of dynamic storage allocation,
in  which each bandwidth request must be accommodated by  a single,
sufficiently  large gap  of available  spectrum.   However, the  process does  correspond
closely  to   the  linked-list   model  of  dynamic   storage  allocation,  in   which  an
available-space list  is maintained and  files can be  fragmented in accordance  with this
list.  This paper also yields a novel  stochastic analysis of such systems.  Note that the
assumption that there is always another request in queue models a system operating at capacity, where the departure rate,
which is equal to the admission rate, is often called maximum throughput.

\subsection*{Notation, State Space, and Probability Model}

We denote by $U$ the size of the request waiting to be allocated bandwidth, and by $U_i$ for $i \geq 1$ the size of the $i$th request behind it. Except in the proof of Theorem~\ref{thm:stab} in Section~\ref{sec:stab}, we omit the dependency in time of these variables for the sake of notation.
%
As mentioned in Section~\ref{sec:model}, we denote by $R(t)$ the number of channels (requests that are allocated bandwidth) at time $t$. $F(t)$ and $G(t)$ denote the number of fragments and gaps at time $t$.  A
Markovian description of  this system is clearly more complicated  than the vector $(G(t),
F(t),  R(t))$.  Hence, we now define the state space of the fragmentation process. At a given state, we denote by $r$ the number of channels (requests to which bandwidth has been allocated) and by $s_i$ ($1 \leq i \leq r$) the amount of bandwidth allocated to these requests. A state of the spectrum $[0,1]$ carries  the information given by a sequence in which gaps
alternate with  sets of  contiguous fragments:
\begin{defi}
A state $x$ in the state space ${\cal S}$ of the fragmentation process is denoted by
\begin{equation*}
\label{eq-state}
 x = (L_1, \ldots ,L_r; u)
\end{equation*}
where $u$ is the size  of (amount of bandwidth required by) the request waiting at the head of the queue and $L_i,$ is the list of open subintervals of $[0,1]$
occupied by  the fragments of the  $i$-th channel.  For $x$ to  be admissible, the
open intervals  in $\cup_i  L_i$ must  be mutually disjoint,  and, since  the size  of $u$
exceeds the bandwidth available, $u > 1-\sum_i s_i$ has to hold. 
\end{defi}
Note that for a specific state, the size of the requests in the queue are denoted by $u$ and $u_j, j\geq 1$ and the size of the channels (requests that are already allocated bandwidth) are denoted by $s_i, 1 \leq i \leq r$.\footnote{For simplicity of the presentation, we did not use the notation $s_i$ in the example provided in Figure~\ref{Example1}.}

Let $(X(t))$ be the process living in the state space $\mathcal{S}$. In the \emph{probability model} used in this paper, the bandwidth requests ($U_i$) are  independent random variables which are uniformly distributed  on $(0,\alpha], 0 < \alpha \leq 1$, and request residence times are i.i.d.\ mean-$1$ exponentials.\footnote{The assumption that request sizes are uniformly distributed is for convenience. The results hold for more general distributions.} Under this probability model, $(X(t))$ is a Markov process on $\mathcal{S}$.

\section{Experimental Results}\label{sec:expr}

In this section we describe an experimental study of the model described above. This serves
two related roles. First, it brings out characteristics of the
fragmentation process that need to be borne in mind in
implementations, particularly where these characteristics show
conditions (e.g., parameter settings) that must be avoided, if a
system with fragmentation is to operate efficiently. The second role
is that of experimental mathematics, in which results indicate where
behavior might well be formalized and rigorously proved as a
contribution to mathematical foundations.  In the latter role, this
section leads up to the next two sections, which formalize the stability
of the  fragmentation process.


The experiments were conducted with a discrete-event
simulator written in C. In general,
the tool is capable of simulating stochastic request
arrival/departure processes. For this paper, however, the arrival process
was effectively inoperative, since the interest here is behavior while the
system is operating at capacity (i.e., at maximum throughput). Admissions are
made whenever the waiting request fits into available spectrum, and once made
the waiting request is immediately replaced by another with independent samples
from the  required-bandwidth and residence-time distributions.
The admission process continues in this way,
effectively simulating a queue that never empties, until no
further admissions can be made and one or more departures need to occur.
The excellent accuracy of the tool was established in tests
against exact queueing results, and exact
results from \cite{KipnisR1990}.  The verification details are omitted and can be found in Appendix \ref{app:validation}.

Recall that bandwidth requests are uniformly distributed on $(0,\alpha], 0 < \alpha \leq 1$. Hence, the principal parameter of the experimental model will be $\alpha$. The simulations of stationary behavior were most demanding, of
course, for small $\alpha$. For every $\alpha$ value, $0.01 \leq \alpha \leq 1$, $20$
million departure events were simulated starting in an empty state,
with data collected for the last $10$ million events. For $0.001 \leq \alpha
\leq 0.01$, $100$ million departure events were processed and data
collection was performed during the last $50$ million events.

\begin{figure}[!t]
\begin{center}
\scalebox{0.4}{ \includegraphics{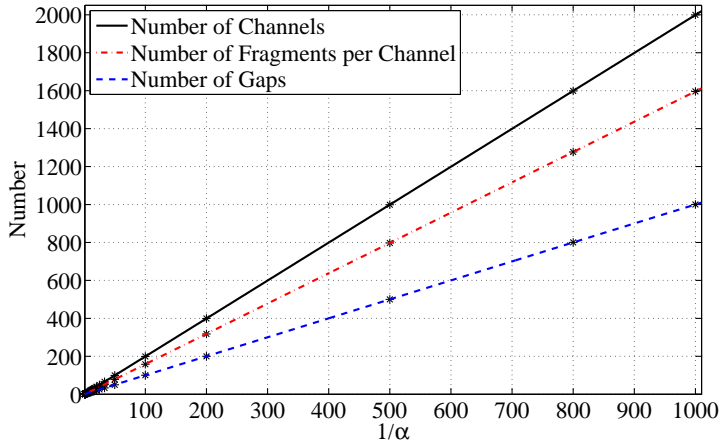}}
 \caption{Average numbers of channels, of gaps, and of fragments per
 channel.}
 \label{fig-linear-frag}
\end{center}
\end{figure}

The results for
the average number of channels, the average number of gaps, and the
average number of fragments per channel are shown in Figure~\ref{fig-linear-frag}. %
The curves are nearly linear in $1/\alpha$ (the errors in the linear fits are within the thickness of the printed
lines). In particular, the asymptotic
average number of channels in the spectrum is $2/\alpha$. 
When requests are large relative to the spectrum
(i.e., for $\alpha >1/3$), the behavior is not given by functions
quite so simple. As such cases are of less practical
interest, we omit the relevant data due to space
constraints.

\begin{figure}[t]
\begin{center}
\scalebox{0.4}{ \includegraphics{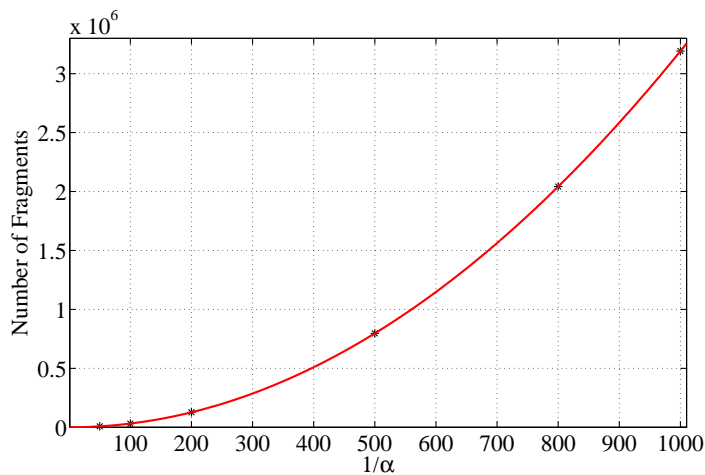}}
 \caption{Average total number of fragments vs. $1/\alpha$: a quadratic fit.}
 \label{fig-quad-frag}
\end{center}
\end{figure}
The asymptotic linear growth of the average number of channels as a
function of channel size is obvious, but the linearity of the other
two measures is not so obvious.  A closer look shows that the
average number of gaps is almost exactly one half the average number
of channels for even relatively small $1/\alpha$. This is an
unexpected version of Knuth's 50\% rule for dynamic storage
allocation.  We return to this behavior in the next section, where
we prove a 50\% limit law.
The linear growth of the average number of fragments
per channel may also be unexpected at
first glance:  the fragmentation of channels  {\it increases}
as the  average channel size
{\it decreases}.   This linear growth  implies the
quadratic  growth of the  average total
number of fragments plotted in Figure \ref{fig-quad-frag}
(the accuracy of the fit is as
before: the  error is within the  thickness of the  printed lines).

\begin{figure}[!t]
\begin{center}
\scalebox{0.4}{ \includegraphics{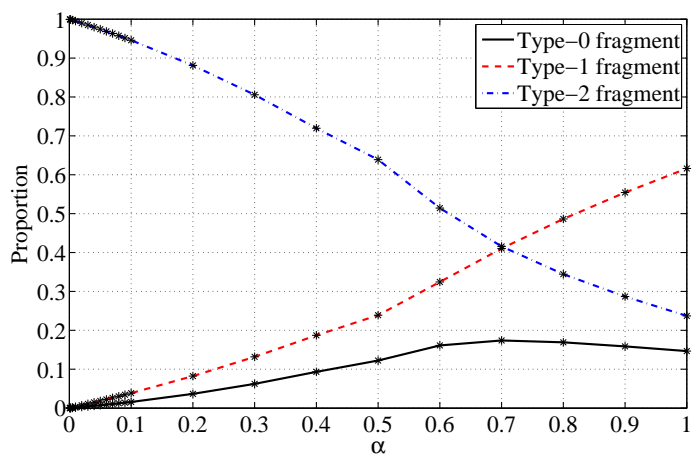}}
 \caption{Percentage of type-$i$ fragments.}
 \label{fig:frag-type}
\end{center}
\end{figure}

The analysis in the later
sections  will focus  largely  on tracking  fragment  {\em types}  defined  as follows:  a
fragment  is of  type-$i$, if  it is  adjacent to  $0,1,~{\rm or}~2$  fragments. It can be seen in Figure \ref{fig:frag-type} that for small $\alpha$, more than 90\% of the fragments are type-$2$ fragments. In addition, clearly, the number of type-$0$ and type-$1$ fragments is a function of the number of gaps. These observations and the  results illustrated in
Figures \ref{fig-linear-frag} and \ref{fig-quad-frag} indicate that, even for relatively
small $1/\alpha$, the average total number  of type-$0$ and type-$1$ fragments grows linearly in
$1/\alpha$, but the average number of type-$2$ fragments grows quadratically.

\begin{figure}[!t]
\begin{center}
 \subfigure[$0 \leq \alpha \leq 1$]{\label{fig-size-frag-gap-a}
\scalebox{0.4}{\includegraphics{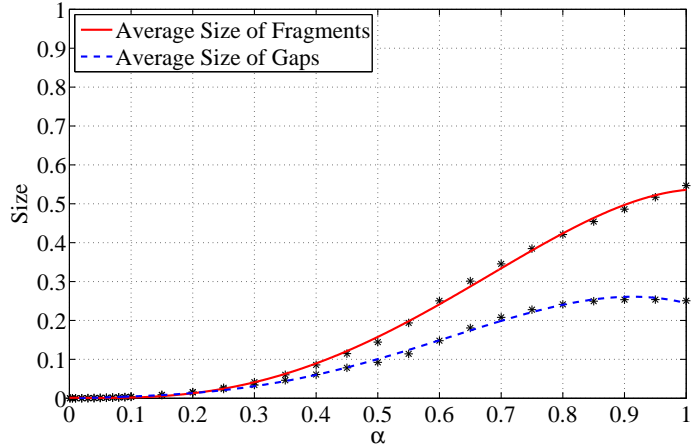}}}
 \subfigure[$0 \leq \alpha \leq 0.1$]{\label{fig-size-frag-gap-b}
\scalebox{0.4}{ \includegraphics{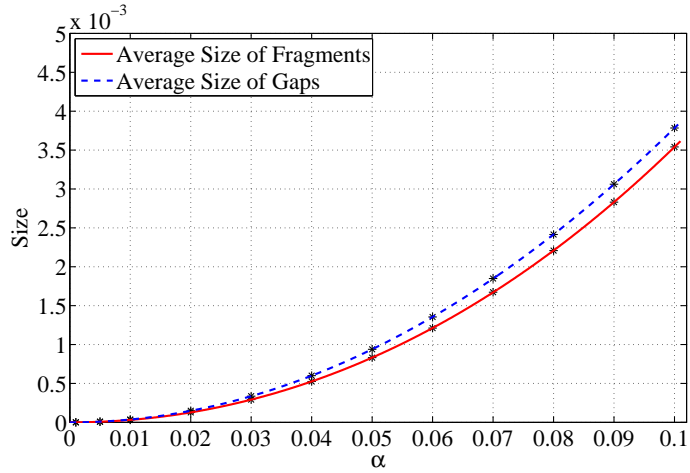}}}
 \vspace*{-0.5cm}
 \caption{Average sizes of fragments and gaps.}
 \label{fig-size-frag-gap}
\end{center}
\end{figure}
Figure \ref{fig-size-frag-gap}  compares the  average gap and fragment sizes.  As might be expected, for relatively  small $\alpha$, they are close to each other.
The relation holds even for moderately  large $\alpha$, although for
$\alpha$ rather  close to  1, the difference  amounts to  about a factor  of 2.  With this
property and the 50\% rule suggested by Figure \ref{fig-linear-frag}, the linear growth in
the number of fragments per channel (shown in Figure \ref{fig-linear-frag}) is easily explained for moderately small $\alpha$ in the following way.

As mentioned above, for moderately small  $\alpha$, the number of channels is approximately $2/\alpha$ (i.e., the spectrum size divided by the average request
size). Due to the 50\% rule, the number  of gaps is roughly $1/\alpha$. At any time, the
total size of the gaps  is at most $\alpha$, since  there is a request waiting  for departures whose
requested bandwidth exceeds that total size of gaps. Therefore, at  most $\alpha$ available bandwidth is spread  among $1/\alpha$ gaps,
giving an average gap  size on the order of $\alpha^2$. The average fragment size is
at most (and indeed very close to) the average gap size. The fragments must occupy at least $1-\alpha$ of the spectrum (since, as mentioned above, at
most a fraction  $\alpha$ of the spectrum is devoted to gaps).  Thus, the  number of fragments
must be on the  order of $1/\alpha^2$, and so the average  number of fragments per channel
must be on  the order of (in  particular, linear in) $1/\alpha$. As  $\alpha \rightarrow 0$, the  asymptotics of these
estimates become more precise.

%

\section{Numbers of Fragments and Gaps}\label{sec:behav}
As mentioned in Section~\ref{sec:model}, under our probability model, $(X(t))$ (the process living in the state space $\Scal$)  is a Markov
process on $\Scal$. In this section, we obtain analytical results regarding the number of fragments and gaps under this process. We begin by formally defining the fragment types which were discussed in Section~\ref{sec:expr}.
%
\begin{defi}
For  $i =$ 0,1, or 2, a fragment is said to be of type~$i$ if it touches
exactly $i$ other fragments.  $N_i(t)$ denotes the number of type~$i$ fragments at time
$t$, so that $F(t) = N_0(t)+N_1(t)+N_2(t)$ is the total number of fragments.
\end{defi}
Let $\sigma(t)$ denotes the sum of the number of fragments and gaps,
\begin{equation}\label{def-sigma}
\sigma(t)=F(t)+G(t).
\end{equation}
The number $G(t)$ of gaps and the numbers of fragment types are related as
follows.
\begin{lemma} \label{lemma:comb}
    With probability $1$,
    \begin{equation} \label{eq:G}
        G(t) = N_0(t) + \frac{1}{2} N_1(t) + I(t)
    \end{equation}
For any $t \geq 0$, where $I(t)=1$, if there is a gap starting at the
origin, and 0 otherwise.
\end{lemma}
\begin{proof} Each gap, except for boundary
gaps starting at 0 or ending at 1, separates two fragments.   Two
gaps surround a type-0 fragment not touching the origin, only one
touches a type-1 fragment not touching the origin, and none touch a
type-2 fragment, so the gaps strictly inside $(0,1)$ are double-counted in $2N_0(t)
+ N_1(t)$. Gaps at the  boundaries are counted only once in this
expression, so if there are gaps touching each boundary, 2 must be
added to $2N_0(t) + N_1(t)$ to produce a double count of all gaps.
Then $N_0 (t) + N_1(t)/2 + 1$ counts the gaps as called for by the
lemma.

With probability $1$, a gap always touches the boundary at $1$, so the only
case left to consider is the absence of a gap touching the origin.
In this case, there is a type-0 or type-1 fragment touching the
origin, and so a nonexistent gap has been counted in $2N_0(t)+N_1(t)$.
This over-count cancels the under-count of the gap touching 1, and
so no correction term is needed, i.e., $N_0(t) + N_1(t)/2$ counts
all gaps as stated in the lemma.
\end{proof}
\begin{defi}
Let $(t_k)$ denote the sequence of departure times, let $D_i(t_k)$
denote the number of type~$i$ fragments in the channel leaving at
time $t_k$, and let $A(t_k)$ denote the number of requests admitted
to the spectrum at time $t_k$.  Finally, define the {\em
drift} in the total number of fragments and gaps:
\begin{equation}
\label{fragplusgap} \Delta \sigma(t_k) \stackrel{\text{def}}{=}
\sigma(t_k) - \sigma(t_{k-1}).
\end{equation}
\end{defi}
The following lemma is the basis of the stability analysis
of~$\sigma(t)$ in Section \ref{sec:stab}, and the 50\% rule proved later in
this section.
\begin{lemma}
\label{main} With probability~$1$, the departure at $t_k$ creates the
following change in the total number of fragments and gaps:
\begin{equation} \label{main-eq}
\Delta \sigma (t_k)   = A(t_k) - 2D_0(t_k) - D_1(t_k) + J(t_k), ~~k
\geq 1
\end{equation}
with $t_0=0$, and  $J(t_k) = 1$, if a fragment starting at the origin is
in the departing channel, and 0 otherwise.
\end{lemma}
\begin{proof}
With probability 1, each new channel allocation covers completely
every gap it is allocated, except for the last one, which is only
partially covered. Thus, with probability 1, each new channel
allocation changes gaps to fragments, except for the last gap which
is changed to a fragment plus a gap; this adds one to
$\sigma(t_{k-1})$ for each admission,
which accounts for the total of $A(t_k)$ in (\ref{main-eq}).

Two fragments of the same channel can not be contiguous, so it is
correct to add up the changes created by departing fragments, with
each being treated separately.  Suppose first that there is no
fragment $(0,b)$ against the origin. Then for every type-0 fragment in
the departing channel, two gaps and a fragment are replaced by a
single gap for a net decrease of two, and for every departing type-1
fragment, a gap and a fragment are replaced by a single gap for a
net reduction of one. This gives the reduction of $2D_0(t_k)+
D_1(t_k)$ appearing in (\ref{main-eq}).  If there is a fragment
$(0,b)$, it must be of type 0 or 1; if it is of type 0, then its
departure gives a decrease of one; if it is of type 1, its departure
has no effect.  Each of these contributions is one less than it
would be were the fragment not touching the origin.  There can only
be one such fragment, so the correction shown in $J(t_k)$ for a
fragment $(0,b)$ follows.
\end{proof}

We will denote by $G^-(t_k)$  the total number of
gaps just after the $k$-th departure, but before new admissions, if
any, are made. Note that if we remove $A(t_k)$ from the right-hand
side of (\ref{main-eq}) and add back the total number of departing
fragments at $t_k$, i.e., $D_0(t_k)+D_1(t_k)+D_2(t_k),$ we get the
number of gaps available to admissions at the $k$-th departure:
 \begin{equation} \label{eq:Delta-H}
        G^-(t_k) = G(t_{k-1}) -D_0(t_k) + D_2(t_k) + J(t_k)
    \end{equation}
with $J(t_k) = 1$, if there is a departing fragment $(0,b)$, and $0$
otherwise.

Knuth's widely known \textit{50\% rule} appears in a very different
context than the model here, so it is difficult to anticipate the
apparent fact that it also holds for our fragmentation model.
However, one can argue a similar result assuming only that the
fragmentation process has a stationary distribution.  The result is given
below as an expected value of a ratio, rather than a ratio of expected
values.
%
\begin{theorem}\label{stab-Har}
Assume that the fragmentation process at departure epochs $t_k$ has the
stationary distribution $\pi_\alpha$, then as $\alpha \rightarrow 0,$
$$ \mathbb{E}_{\pi_\alpha}\left( {G(t_k) \over R(t_k)} \right) \sim \frac{1}{2} $$
\end{theorem}
%
\begin{proof}
In the stationary regime, one has, by Lemma~\ref{main},
\begin{align*}
\mathbb{E}_{\pi_\alpha} [ \sigma(t_{k}) - \sigma(t_{k-1})] &= \\
\mathbb{E}_{\pi_\alpha}[A(t_{k}) - 2D_0(t_{k}) -
D_1(t_{k}) + J(t_{k})]
& =  0,
\end{align*}
and to balance departure and admission rates, we must have
$\mathbb{E}_{\pi_\alpha} A(t_{k}) = 1$.  Thus, we can write
\begin{equation}
\label{whatever}
 \mathbb{E}_{\pi_\alpha}[2D_0(t_{k}) + D_1(t_{k}) - J(t_{k})] = 1.
\end{equation}
Now for a state $x$ at time $t_k$ having $r$ channels and $N_i(t_k)$
type-$i$ ($i=0,1,2$) fragments, we have from Lemma \ref{lemma:comb}
\begin{align*}
\mathbb{E}_x[2D_0(t_{k}) + D_1(t_{k}) - J(t_{k})] & = 
(2N_0(t_k) + N_1(t_k))/r - \mathbb{E}_x[J(t_{k})]  \\ & =
2[G(t_k)  - I(t_k)]/ r - \mathbb{E}_x[J(t_{k})]  =   2G(t_k) / r + O\left(1/r\right)
\end{align*}
Thus, dropping the $O(1/r)$  term that tends to 0 with $\alpha$ uniformly in $x$, the
expectation in (\ref{whatever}) proves the theorem.
\end{proof}

\section{Stability Results} \label{sec:stab}
This section  establishes that the average total number of fragments
and gaps remains bounded and that, for certain distributions of
request sizes, ergodicity holds. The analysis leads to the
two following results. Recall that $(t_n)$ is the sequence of
departure times ($t_0 = 0$) and $\sigma(t) = F(t)+G(t)$, defined
in~\eqref{def-sigma}, is the total number of fragments and gaps.
\begin{theorem} \label{thm:sup}
There exists some $\eta>0$ such that, for any initial state $x \in \Scal$,
 \begin{equation} \label{eq:sup-n}
    \sup_{n \geq 1} \E_x \left( e^{\eta \sigma(t_n)} \right) < +\infty.
    \end{equation}
\end{theorem}
Clearly, this implies that for any initial state $x \in \Scal$, the
sequence $(\E_x(\sigma(t_n)), n \geq 0)$ is bounded. With an
additional assumption on the distribution of the request size, a stronger stability result can be proved.
\begin{theorem} \label{thm:stab}
When $\alpha  > 1/2$, the process $(X(t))$ is positive Harris
recurrent; in particular, it has a unique stationary distribution.
\end{theorem}
A criterion for finite exponential moments using a Lyapunov function
is established next.  Then, we provide  some estimates of the drift
of the number of fragments between departures which will show us how
to construct a Lyapunov function.  After constructing this function,
we will be in position to prove  the boundedness of exponential
moments.

\subsection*{A Criterion for Finite Exponential Moments}
Before stating the main result, some  results on Markov chains are
needed. In the sequel, $\leq_{st}$ refers to stochastic ordering,
i.e., $V \leq_{st} Z$ means that $\E(f(V)) \leq \E(f(Z))$ for any
increasing function $f$. For reasons that will become clear in Lemma~\ref{lemma:Delta},
the following lemma focuses on admissions at $4$ consecutive departure
times. Recall that $(U_i, i \geq 1)$ are the sizes of the requests waiting to be allocated bandwidth (after the first one $U$), which are assumed to i.i.d.
\begin{lemma}\label{lemma:coupling}
    The random variable $A(t_1) + \cdots + A(t_4)$ is stochastically dominated by a random
    variable $Z$ such that $\E(e^{\lambda Z}) < +\infty$ for some $\lambda > 0$.
\end{lemma}
\begin{proof}
It is clear that $A(t_1) \leq Z+1$ where
    \[ Z = 1 + \inf \{ n \geq 1: U_1 + \cdots + U_n \geq 1\}. \]
The Markov inequality shows that for any $z
\geq 0$,
 \[ \P(Z \geq z+1) = \P(U_1 + \cdots + U_z \leq 1) \leq
    e \left( \E\left( e^{-U_1} \right) \right)^z \]
and so $\E(e^{\eta Z})$ is finite for $\eta > 0$ small enough. From
this observation, it is not difficult to extend the result to
$A(t_1) + \cdots + A(t_4)$ instead of just
    $A(t_1)$.
\end{proof}
This lemma shows in particular that
\[  \xi \stackrel{\text{def}}{=} \sup_{i \geq 1}
\sup_{x \in \Scal} \E_x(A(t_i)) < +\infty. \] is well-defined; this
constant will be used repeatedly throughout the rest of the
analysis. The proof of the following lemma is standard, and therefore,
omitted.
\begin{lemma} \label{lemma}
Let $Z \geq 0$ be a positive, real-valued random variable such that
$\E(e^{\lambda Z}) < +\infty$ for some $\lambda > 0$, and define
    $c = \lambda^{-2} \E(e^{\lambda Z} - 1 - \lambda Z). $
Then for any $0 \leq \varepsilon \leq \lambda$ and any real-valued
random variable $V$ such that $V \leq_{\text{st}} Z$, we have
    $ \E \left( e^{\varepsilon V} \right) \leq 1 + \varepsilon \E(V) + \varepsilon^2 c.
    $
\end{lemma}
The following result is closely related to result of
Hajek~\cite{Hajek82:0}. The proof can be found in Appendix \ref{app:proof}.
\begin{theorem} \label{thm:hajek}
Let $(Y_k)$ be a discrete-time, continuous state-space Markov chain
such that for some function $f \geq 0$, there exist $K, \gamma > 0$
such that for any initial state $y$ with $f(y) > K$,
    $ \E_y(f(Y_1) - f(Y_0)) \leq -\gamma. $
Assume that there exists a random variable $Z$ such that for any
initial state $y$, $Z$ dominates stochastically the random variable
$f(Y_1) - f(Y_0)$ under $\P_y$. Assume finally that $\E(e^{\lambda
Z}) < +\infty$ for some $\lambda > 0$. Then there exist $\eta > 0$
and $0 \leq C < +\infty$ such that for any initial state $y$,
\[ \sup_{n \geq 1} \E_y \left( e^{\eta f(Y_n)} \right) \leq e^{\eta f(y)} + C. \]
\end{theorem}
Theorem~\ref{thm:hajek} will be applied to the Markov chain
$(X(t_{4n}))$ with a function $f$ of the form $\sigma_\kappa =
\sigma + \kappa r$ for some $\kappa
> 0$ suitably chosen. $(X(t_{4n}))$ is not the most natural choice
at first glance, but it appears to be needed because of the
complexity of the state space.

It is clear that $\sigma(t_1) - \sigma(t_0) \leq  A(t_1) + 1$, so
that \[ \sigma_\kappa(t_4) - \sigma_\kappa(t_0) \leq (\kappa + 1)
(A(t_1) + \cdots + A(t_4)) + 4
\] and therefore, by Lemma~\ref{lemma:coupling},
$\sigma_\kappa(t_4)-\sigma_\kappa(t_0)$ is stochastically dominated
by some random variable $Z$ with an exponential moment. Therefore,
one has to establish a negative drift relation for
$\sigma_\kappa(t_4) - \sigma_\kappa(t_0)$. This is the purpose of
the following two subsections.

\subsection*{Evolution of the Number of Fragments}
Recall that  $x\in\Scal$, the initial state of
the system, has $r$ active channels, and define the total available gap size
$h  = 1  - (s_1  {+} \cdots  {+} s_r).$  Time $0$ referring to
the initial state $x$  will usually be omitted; e.g., $\sigma(0),
F(0), G(0), \ldots$ will be  simplified to $\sigma, F, G, \ldots$.
Recall  that $\Delta \sigma(t_n) $ is defined in~\eqref{fragplusgap} as $\sigma(t_n) -
\sigma(t_{n-1}).$
%
\begin{lemma} \label{lemma:Delta}
Fix $0 < \varepsilon < 1$ and $0 < \eta < 1/2$, and  let $x \in
\Scal$ be an initial state such that $\sigma = G+F \geq 2K + 1$ for
some fixed $K \geq 0$.

Then $F = N_0 + N_1 + N_2 \geq K$,
and\renewcommand{\labelenumi}{\arabic{enumi})}
\begin{enumerate}
    \item If $r = 1$, then $\E_x(\Delta \sigma(t_1)) \leq \xi - K$.
\end{enumerate}

\begin{enumerate}\setcounter{enumi}{1}
    \item If $r>1$ and $N_0 + N_1 \geq \varepsilon K$, then
    $$ \E_x(\Delta\sigma(t_1)) \leq \xi + 1 - \frac{\varepsilon
    K}{r}.$$
\end{enumerate}
Assume in the remaining cases that $r > 1$, define $K' =
K\big((1-\varepsilon)/ r - \varepsilon\big)^+$, and let $i^* \in
\{1, \ldots, r\}$ index a channel $L_{i^*}$ in $x$ with the most
type-$2$ fragments.
    \begin{enumerate}\setcounter{enumi}{2}
        \item
If $N_0 + N_1 \leq \varepsilon K$ and $u > h + s_{i^*}$, then
        \begin{equation} \label{eq:case-N3-1}
        \E_x(\Delta\sigma(t_2)) \leq \xi + 2 - \frac{K'}{r(r-1)}.
        \end{equation}
        \item If $N_0 + N_1 \leq \varepsilon K$, $u < h + s_{i^*}$ and $h + s_{i^*} < \eta \alpha$,
        then
        \begin{equation} \label{eq:case-N3-2}
            \E_x(\Delta \sigma(t_3)) \leq \xi + 2 -
\frac{(1-\eta)K'}{r^2 (r - 1)}.
        \end{equation}
        \item If $N_0 + N_1 \leq \varepsilon K$, $u < h + s_{i^*}$ and $\eta \alpha
        < h + s_{i^*}$, then there exists a $\gamma(\eta) > 0$ such that
        \begin{equation} \label{eq:case-N3-3}
            \E_x(\Delta \sigma(t_4)) \leq \xi + 2 - \frac{\gamma(\eta) K'}{r^5}.
        \end{equation}
    \end{enumerate}
    It follows that there exists a $\overline \xi > 0$ and a function $\psi(r) > 0$
    such that for any $x$ with $\sigma \geq 2K+1$,
    \begin{equation} \label{eq:all}
        \E_x(\sigma(t_4) - \sigma) \leq \overline \xi - K \psi(r).
    \end{equation}
\end{lemma}
\begin{proof}
As is readily verified, $G \leq F+1$, so $2K+1 \leq \sigma = F + G
\leq 2F + 1$, and hence $F \geq K$ as claimed. In what follows, we
use repeatedly the two following simple facts:
\begin{equation}
 \label{sf:2}
\E_x \left( D_0(t_1) + D_1(t_1) \right) = (N_0 + N_1) / r,
\end{equation}
and by Lemma \ref{lemma:comb},
\begin{equation}
\label{sf:1} G \geq K \Rightarrow N_0 + N_1 \geq K - 1.
\end{equation}

\noindent \textit{{--- First case: $r = 1$.}} Then, right after the
only channel initially present leaves, there is no channel allocated bandwidth, and
therefore, $\sigma(t_1) = A(t_1)$. Note that $r = 1$ is only possible
when $\alpha > 1/2$, and in this case the possibility for a channel
to be alone is crucial in the proof of the Harris recurrence
stated in Theorem~\ref{thm:stab}.

\noindent \textit{{--- Second case: $r > 1$, $N_0 + N_1 \geq \varepsilon
K$.}} Then the inequality follows from~\eqref{main-eq}:
\[
    \E_x(\Delta \sigma(t_1)) \leq \xi + 1 - \E_x(D_0(t_1) + D_1(t_1))
    = \xi + 1 - \frac{N_0 + N_1}{r} \leq \xi + 1 - \frac{\varepsilon K}{r}.
\]

In the 3 remaining cases, let $N^*_j$ denote the number of type-$j$
fragments in any channel $i^*$ which has the most type-$2$
fragments. If $N_0 + N_1 \leq \varepsilon K$, then since $F \geq K$,
necessarily $N_2 \geq (1-\varepsilon) K$ and $N^*_2 \geq
(1-\varepsilon)K/r.$
  Define the event $D^* = \{ \text{channel } L_{i^*} \text{
leaves at } t_1 \}$ and recall that $G^-$ denotes the number of gaps
right after $L_{i^*}$ leaves but before new admissions, if any, are
made. It follows from~\eqref{eq:Delta-H} that $G^- \geq K'$ in the
event $D^*$, since
    \[ G^- = G - N^*_0 + N^*_1 + J(t_1)
    \geq \left (- \varepsilon K + (1-\varepsilon)K / r \right)^+ = K'. \]
The remaining analysis tacitly assumes that $r>1$, that $N_0 + N_1 \leq
\varepsilon K$ and that the channel $L_{i^*}$ leaves at $t_1$.

\noindent \textit{{--- Third case: $u > h + s_{i^*}$.}} Then $A(t_1)
= 0$, since when~$L_{i^*}$ leaves it does not provide enough
additional bandwidth for~$U$. In particular, $R(t_1) = r-1$ and
$G(t_1) = G^- \geq K'$, and so
    \[
    \E_x(\Delta \sigma(t_2))
    \leq \xi + 1 - \E_x(D_0(t_2) + D_1(t_2); D^*).
    \]
The strong Markov property makes it possible to lower-bound this
last term.
\begin{multline*}
\E_x(D_0(t_2) + D_1(t_2); D^*) = \E_x(\E_{X(t_1)}(D_0(t_1) + D_1(t_1)) ; D^*)\\
= \E_x \left(\frac{(N_1 + N_2)(t_1)}{R(t_1)} ; D^* \right) \geq
\frac{K' - 1}{r - 1} \P_x (D^*) = \frac{K' - 1}{r(r - 1)}
\end{multline*}
and therefore, $\E_x(\Delta \sigma(t_2)) \leq \xi + 2 - K' /
(r(r-1))$.

    \noindent \textit{{--- Fourth case: $u < h + s_{i^*} < \eta \alpha$.}}
    In this case $U$ is admitted at $t_1$. Thus it makes sense to define the
    event
    \[
        E_4 = D^* \, \cap \, \{ U \text{ leaves at } t_2 \text{ and } U_1 > \eta \alpha \}.
    \]
Then as before
    \[ \E_x(\Delta \sigma(t_3)) \leq \xi + 1 - \E_x(D_0(t_3) + D_1(t_3) ; E_4). \]
In the event $E_4$, $U$ is admitted at $t_1$ and leaves at
$t_2$, while $U_1$ stays blocked at $t_1$ and $t_2$,
so that $G(t_2) = G^- \geq K'$ and $R(t_2) = r - 1$. Hence as in the
second case,
\[ \E_x(D_0(t_3) + D_1(t_3) ; E_4) \geq \frac{K' - 1}{r - 1} \P_x (E_4)
\geq \frac{(1-\eta)K'}{r^2(r-1)} - 1 \] since $\P_x(E_4) =
(1-\eta)/r^2$. Thus~\eqref{eq:case-N3-2} holds.

\noindent \textit{{--- Fifth case: $u < h + s_{i^*}$ {\rm and} $\eta
\alpha < h + s_{i^*}$.}} Again, $U$ is admitted at~$t_1$. Letting
$U_i$ denote the sizes of the requests behind $U$, define the event
$$ B = \{U_i < \eta \alpha,\ i = 1, \ldots, \tau \text{ and } U_{\tau
+ 1} > 2 \eta \alpha\} $$ with $\tau = \inf \{ n \geq 0: U_1 +
\cdots + U_n > h + s_{i^*} - \eta \alpha \}$ and
    $ E_5' = D^* \, \cap \, B \, \cap \, \{U \text{ leaves at } t_2 \}.
    $
It is readily verified that $1 \leq \tau < +\infty$ almost surely.
Moreover, one has in $E_5'$
    \[ 0 < h^* \stackrel{\text{def}}{=} h + s_{i^*} - (U_1 + \cdots + U_{\tau})
    < \eta \alpha < U_{\tau + 1}. \]
This means that at $t_2$, exactly $\tau$ new requests $U_1, \ldots,
U_\tau$ have been admitted, and $U_{\tau + 1}$ is blocked. Moreover,
for any $i \in \{1, \ldots, \tau\}$, one has $h^* + U_i < 2 \eta
\alpha < U_{\tau + 1},$ so that if one of the $\tau$ channels
allocated to the $(U_i)$ leaves, $U_{\tau + 1}$ remains blocked.

When $L_{i^*}$ left, there were $G^- \geq K'$ gaps; in the remainder
of the analysis, we call an \emph{initial gap} a gap present right
after $L_{i^*}$ left. After $L_{i^*}$ left, $U$ and $A(t_1) - 1$ new
requests were admitted, and then $U$ left and $A(t_2)$ new requests
were admitted at $t_2$. Thus, at $t_2$, each initial gap is in
either of two states: either it is completely filled, or it is still
a gap, i.e., it has not been filled completely. Let $k$ be the
number of initial gaps completely filled at $t_2$, and let $k' = G^-
- k$: then $k + k' = G^- \geq K'.$ In each initial gap completely
covered at $t_2$, there is at least one type-$2$ fragment of one of
the $\tau$ new channels. Therefore,
    $ N_{1, 2} + N_{2, 2} + \cdots + N_{\tau, 2} \geq k $
with $N_{i,2}$ the number of type-$2$ fragments of the channel corresponding to $U$.
In particular there is a channel $L_{j^*},~ j^* \in \{ 1, \ldots,
\tau \}$ with at least the average $k / \tau$ of type-$2$ fragments:
$N_{j^*, 2} \geq k / \tau$. Define finally the event $E_5 = E_5' \,
\cap \, \{ L_{j^*} \text{ leaves at } t_3 \}.$ Since $h^* + U_{j^*}
< U_{\tau + 1}$, then $U_{\tau + 1}$ remains blocked at $t_3$ when
$E_5$ occurs, and therefore (note that when $j^*$ leaves, some gaps
may merge, but not two initial gaps),
    \[ G(t_3) \geq N_{j^*, 2} + k' \geq k / \tau + k' \geq (k + k') / \tau \geq K' / \tau. \]
Now we proceed as before, to obtain
    \[ \E_x(\Delta \sigma(t_4)) \leq \xi + 1 - \E_x(D_0(t_4) + D_1(t_4) ; E_5) \]
and, using the Markov property at time $t_3$,
\begin{multline*}
    \E_x(D_0(t_4) + D_1(t_4) ; E_5) = \E_x( \E_{X(t_3)} (D_0(t_1) + D_1(t_1)) ; E_5)\\
    = \E_x \left( \frac{(N_0 + N_2)(t_3)}{R(t_3)} ; E_5 \right)
    \geq \E_x \left( \frac{(K' / \tau - 1)^+}{r + \tau - 2} ; E_5 \right)
\end{multline*}
since $R(t_3) = r + \tau - 2$ in $E_5$. The same kind of reasoning
as before then leads to
\[
    \E_x \left( \frac{(K' / \tau - 1)^+}{r + \tau - 2} ; E_5 \right)
    \geq \frac{K'}{r^5} f(\eta, h + s_{i^*} - \alpha \eta) - 1
\]
with the function $f(\eta, \cdot)$ defined for $y > 0$ by
\[ f(\eta, y) = \E \left((1 + \tau(y))^{-5} ; B(\eta, y) \right) \]
with $\tau(y) = \inf\{n \geq 1: U_1 + \cdots + U_n \geq y\}$ and
\[ B(\eta, y) = \{ U_i < \eta \alpha, i = 1, \ldots, \tau(y)
\text{ and } U_{\tau(y)+1} > 2 \eta \alpha \}. \] It is not
difficult to show that
    $ \gamma(\eta) = \inf_{0 < y < 1} f(\eta, y) > 0 $
which then gives the result.

It remains to prove~\eqref{eq:all}. One only needs to assemble the
various bounds, taking into account  that $\E_x(\Delta
\sigma (t_i)) \leq \xi + 1$ for any $x \in \Scal$ and $i \geq 0$, to
arrive at 4 separate bounds on $\E_x(\sigma(t_4)-\sigma)$.  For
example, using the former bound for the first two terms and the last
term of
$\E_x(\sigma(t_4)-\sigma) = \sum_{1 \leq i \leq 4} \E_x \Delta \sigma(t_i)$
and then the bound in~\eqref{eq:case-N3-2} for the third term, we
get that one of the 4 bounds, which applies when $x$ satisfies the
inequalities of the fourth case, is
$$
\E_x(\sigma(t_4)-\sigma) \leq  4\xi + 5 - \frac{(1-\eta)K'}{r^5}
$$
Computing the minimum over these bounds with $\eta=1/4$
and~$\varepsilon=1/r^{2}$, one obtains~\eqref{eq:all} after setting
$\overline{\xi} = 4 \xi +5$ and
    \[ \psi(r) = \frac{ \varphi(r)}{r^6} \times ((1-\eta) \wedge \gamma) \]
with $\varphi(r) = 1-2r^{-2}$. This concludes the proof.
\end{proof}
We turn now to  the case where $r$ is large.  In this case,  the
negative drift comes from the fact that, except  perhaps at $t_1$,
with high probability there  is no admission at a departure, since
the  channel that leaves is small with high  probability. However,
we see in~\eqref{main-eq} that this  is not enough for  $\sigma$ to
decay, for one  would need at least one type-$0$ or type-$1$
fragment to leave as well, and it can be the case that most
fragments are of type-$2$. The second term of the Lyapunov function
allows us to get around this  problem.  Since  the  variation
$\Delta R(t_k)$  in  the number  of  channels at  a departure is
exactly equal to $A(t_k) - 1$, one readily gets that
\[ \sigma_\kappa (t_4) - \sigma_\kappa = (\sigma(t_4) - \sigma) +
\kappa (A(t_1)  + \cdots + A(t_4) -  4). \] In particular,  if $x \in \Scal$  is such that
$\sigma  \geq 2K+1$  and $r  \leq  K_r$, then  (from now  on,  we assume  without loss  of
generality that the function  $\psi$ given by~\eqref{eq:all} in Lemma~\ref{lemma:Delta} is
decreasing)
\[ \E_x(\sigma_\kappa(t_4) - \sigma_\kappa)
\leq \overline{\xi} - K \psi(K_r) + 4 \kappa (\xi - 1)
\] whereas if $r \geq K_r$,
\[ \E_x(\sigma_\kappa(t_4) - \sigma_\kappa) \leq 4 (\xi
 + 1) + \kappa \E_x(r(t_4) - r) \] and so we see
that we only need to control $\E_x(r(t_4) - r)$ for~$r$ large.
\begin{lemma} \label{lemma:ell}
There exist $K_r, \gamma_r > 0$ such that if $x \in \Scal$ is such that $r \geq K_r$, then
 $ \E_x(r(t_4) - r) \leq -\gamma_r. $
\end{lemma}
\begin{proof}
Since the technical difficulty of the proof of this inequality is
similar to that of the above proof, we need only give a sketch of
it. From $s_1 + \cdots s_r = 1-h \leq 1$ one gets $\# \{ i : s_i
\geq \gamma \} \leq 1 / \gamma$, and therefore, $\P_x(s_{i_1} \geq
1/\sqrt r) \leq 1/\sqrt r$ with $L_{i_1}$, $i_1 \in \{1, \ldots,
r\},$ the channel that leaves at $t_1$. Thus, when $r$ is large,
with high probability a small channel leaves.

If $h-u$ is away from $\alpha$, then the event $\{u_1 > h - u + s_{i_1} + \cdots + s_{i_4}
\}$  (with $i_k$ defined  similarly) has  high probability,  and in  this event  $A(t_1) +
\cdots + A(t_4) \leq  1$. If in contrast $h-u$ is large,  then with high probability $U_1$
is admitted and with high probability $h - u - U_1$ is away from $\alpha$; hence we can do
the  same again,  and get  that, with  high probability,  $A(t_1) +  \cdots +  A(t_4) \leq
2$. The lemma is proved.
\end{proof}

\subsection*{Construction of a Lyapunov Function}
For $\kappa
> 0$, one defines, for an initial state $x\in\Scal$ of the system,
$\sigma_\kappa \stackrel{\text{def}}{=} \sigma+ \kappa \, r,$ where
$r$  is the number of channels allocated in $x$ and $\sigma=F+G$ is
the sum of the number of fragments and gaps in $x$.
\begin{proposition} {\em (Lyapunov function inequality)} \label{prop:lyapunov}
There exist $\kappa$ and $K > 0$ such that if $x \in \Scal$ is such
that $\sigma_\kappa \geq K$, then
       $ \E_x(\sigma_\kappa(t_4) - \sigma_\kappa) \leq - 1.$
\end{proposition}
\begin{proof}
   Let $K_r$ and $\gamma_r$ be as in Lemma~\ref{lemma:ell}, and take
    $\kappa$ and $K$ as follows:
\[ \kappa = \frac{4 \xi + 5}{\gamma_r}\ \text{ and }\ K =
\frac{8 \xi + 2\overline{\xi} + 2}{\psi(K_r)} + \kappa K_r + 1.
\]

Assume that $\sigma_\kappa \geq K$. If $r \geq K_r$, then
    \[ \E_x(\sigma_\kappa(t_4) - \sigma_\kappa)
    \leq 4(\xi + 1) - \kappa \gamma_r = -1. \]
Otherwise, $r \leq K_r$, and since $\sigma_\kappa \geq K$, this
necessarily gives $\sigma \geq K - \kappa K_r = 2\hat{K}+1$ with
$\hat{K} = (K - \kappa K_r - 1) / 2$. Thus,
    \[ \E_x(\sigma_\kappa(t_4) - \sigma_\kappa) \leq \overline{\xi} -
    \hat{K} \psi(K_r) + 4 \xi = -1 \]
and the proposition follows.
\end{proof}

\subsection*{Proof of Theorem~\ref{thm:sup}}
Theorem~\ref{thm:hajek} and Lemma~\ref{lemma:coupling} applied to
the Markov chain $(X(t_{4n})$, $n \geq 0),$ and the function
$\sigma_\kappa$ show that for some $\eta > 0$ and some constant $0
\leq C < +\infty$,
\[ \sup_{n \geq 0} \E_x \left( e^{\eta \sigma_\kappa(t_{4n}) } \right) \leq e^{\eta \sigma_\kappa} + C. \]
Then the Markov property gives for any $i \geq 0$
\[ \sup_{n \geq 0} \E_x \left( e^{\eta \sigma_\kappa(t_{4n+i}) } \right) \leq \E_x \left( e^{\eta \sigma_\kappa(t_i)} \right) + C < +\infty \]
from which~\eqref{eq:sup-n} follows readily.

\subsection*{Proof of Theorem~\ref{thm:stab}}  In the following
discussion, no conceptual argument is missing, only some formalism
needed to handle the continuous state space $\Scal$. These details
are routine and left to the interested reader. In the analysis
below, requests are said to be {\em big} if their size exceeds
$1/2$.

We argue that $(X(t))$ visits infinitely often a state in which
there are no fragmented channels, and such that all size
distributions remain the same at all visits. This is enough to show
Harris recurrence; see for instance Asmussen~\cite{Asmussen87:0}.
For this purpose, it is convenient to pick a simple regeneration set
$E \subset \Scal$ in which (i) the spectrum is being used by a big
request, alone and with an unfragmented channel of the form $(0,b)$,
and (ii) the request $U$  waiting at the head of the queue is also
big. Each of these has the conditional request-size distribution
given that its size is larger than $1/2$, i.e., the uniform
distribution on $(1/2,\alpha)$, see~\cite{KipnisR1990}.

To verify that $E$ is visited infinitely often, consider the process
$(R(t), U(t))$ with $R(t)$ the number of requests allocated a
channel at time $t$ and $U(t)$ the size of the request at the head
of the queue; the process $(R(t), U(t))$ is simply the process
$(X(t))$ when the data on fragmentation is ignored. This process is
positive Harris recurrent, as shown in Kipnis and
Robert~\cite{KipnisR1990}. In particular it visits infinitely
often states with $R(t) = 1$ and $U(t) > 1/2$; one can add $U_1(t) >
1/2$ as well (i.e., the first request in line behind the head of the queue request is also big), since
this happens with a geometric probability. Then when the only
channel leaves, the process $(X(t))$ enters $E$, since there is
exactly one channel, it is big, it is necessarily
unfragmented and of the form $(0,b)$, and a big request is waiting
at the head of the queue. Moreover, this argument shows that the
time between visits to~$E$ is integrable, which in turn establishes
positive Harris recurrence. This completes the proof.

\section{Algorithms}\label{sec:algorithms}
Although the focus so far has been on measures of fragmentation as a
function of $\alpha$, algorithmic issues are also of obvious
interest.  For example, more uniform patterns of gaps might be an
advantage. The \emph{Linear Scan} (LS), discussed in the previous sections, tends to push
the gaps towards the end of the spectrum, particularly when the
spectrum is viewed at random times in steady state. Interestingly,
our experiments have shown that, for all $\alpha < 1/3$, the
starting position of the first gap in the spectrum remains very
close to $0.64$.

To uniformize gap locations, an alternative
gap scan resembles the {\em Circular Scan} (CS) sequences
of dynamic storage allocation \cite{Knuth1997}. In our case, CS uses
a circular gap list, in which the successor to the last gap in [0,1]
is the first gap in [0,1]. The scan is still linear, but the
starting gap of the scan moves as follows: if the last fragment of
a channel is placed in gap $g$, then the residual gap of $g$ is the
first gap scanned in constructing the next channel. Clearly,
although CS will tend to uniformize gap sizes as a function of
position, boundary effects will persist so long as the spectrum
itself is not circular, i.e., gaps and fragments are not allowed to
overlap the end of the spectrum, a restriction that would likely be
dictated in practice.

\begin{figure}[!t]
\begin{center}
\scalebox{0.4}{\includegraphics{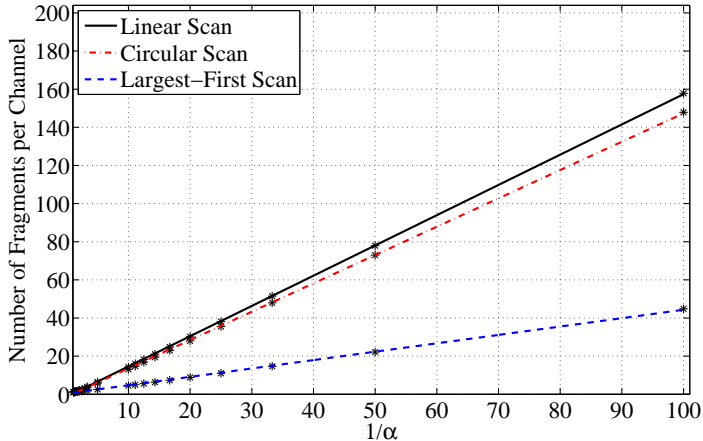}}
 \caption{Average number of fragments per channel.}
 \label{gapsearch1}
\end{center}
\end{figure}
The average number of fragments per channel is a direct measure of
gap-search times, and one that we use here. For values of $\alpha$
expected to be of interest in applications, the effects of CS on
gap-search times are only within a few percent relative to LS,  as
can be seen in Figure~\ref{gapsearch1}.   The figure also shows the average number of fragments per channel for the \emph{Largest-First Scan} (LFS) algorithm. This algorithm is designed to
speed up the process of finding a set of gaps
sufficient to create a new channel. It selects available gaps in a decreasing order of their sizes and allocates them to a request, thereby greedily minimizing the
number of gaps needed to fulfill a request.
The extra mechanism
needed for such a search will of course tend to reduce overall
performance gains.  The results in Figure~\ref{gapsearch1}
for LFS show  a surprisingly large improvement in the
average number of fragments per channel -- as can be seen, a
reduction by a factor more than 3 is achieved by LFS for even
moderately small $\alpha$.

\begin{figure}[!t]
\begin{center}
\scalebox{0.4}{\includegraphics{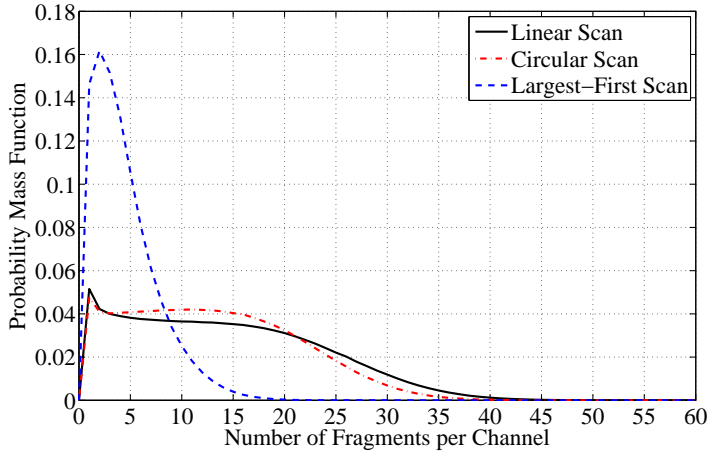}}
 \caption{Distribution of the number of fragments per channel for $\alpha = 0.1$.}
 \label{dist-fragper}
\end{center}
\end{figure}

\begin{figure}[!t]
\begin{center}
\scalebox{0.4}{ \includegraphics{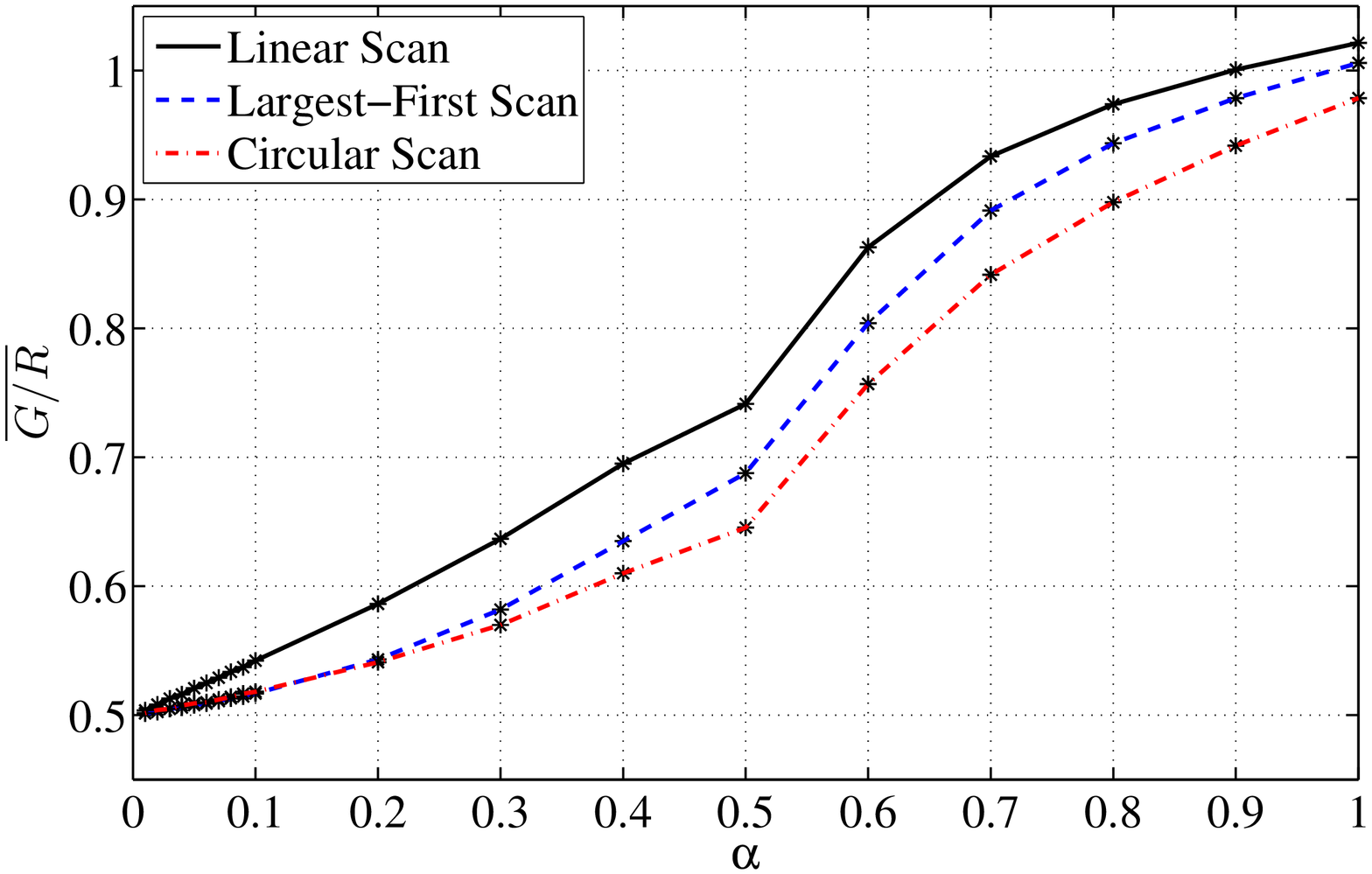}}
 \caption{ $\overline{G/R} \rightarrow 1/2$
 as $\alpha \rightarrow 0$ under LS, LFS, and CS.}
 \label{fifty-law-ls-cs}
\end{center}
\end{figure}
The Probability Mass Functions (pmf's) of the number of fragments per channel are shown in Figure \ref{dist-fragper}.
Notice that while all probabilities are small under LS and CS, the largest applies to the case of no fragmentation at all.  The much more peaked distribution for LFS has both  much smaller mean and variance:  The standard deviation under CS and LS is approximately $1.5$ to $2.0$ times that under LFS.
The same limiting behavior called for by the 50\% law holds for CS
and LFS, which was to be expected, as the arguments supporting the
50\% law did not depend on the sequence in which gaps were scanned.
But an interesting result of our experiments with CS is that the
50\% approximation to expectation of the ratios  is within a couple
of percent {\em even when the maximum request size is as much as 0.2
the spectrum size}. This is easily seen in Figure
\ref{fifty-law-ls-cs}.  The convergence rate of LFS to 50\% is intermediate between LS and CS.
\section{Normal Approximations}\label{sec:normal}

\begin{figure}[!t]
\begin{center}
  \subfigure[$\alpha = 0.5$]{\label{frag0500}
\scalebox{0.4}{ \includegraphics{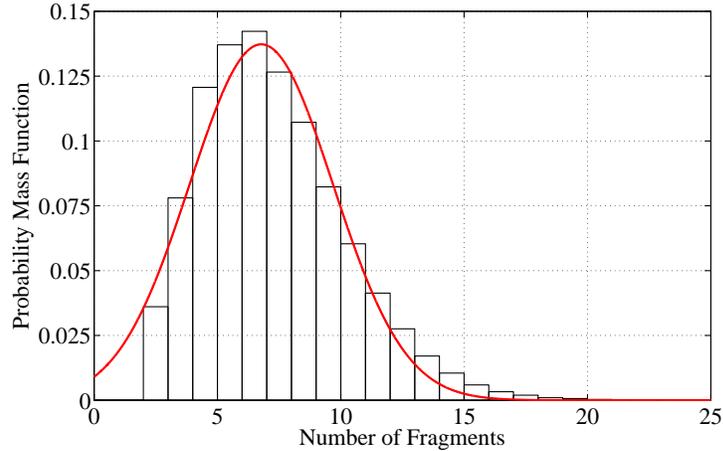}}}
  \subfigure[$\alpha = 0.1$]{\label{frag0100}
\scalebox{0.4}{  \includegraphics{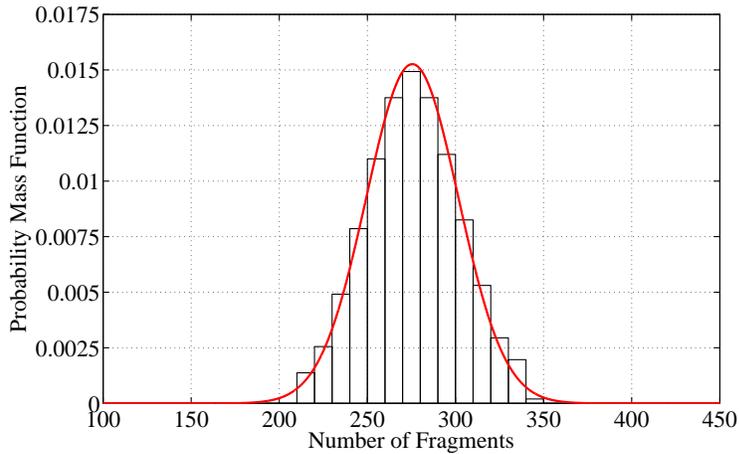}}}
  \caption{Distributions of the total number of fragments and corresponding Normal fits.}
  \label{normal-fits-num-frags}
\end{center}
\end{figure}
After the usual scaling (i.e.,
first centering then normalizing by the standard deviation), the
scaled version of the number of channels, $R$, tends in
probability to the standard Normal, ${\mathcal{N}}(0,1),$ as $M
\rightarrow \infty$, where $M = \lfloor 1/\alpha \rfloor$ (it is convenient to express the asymptotics in this section in terms
of $M$). This result follows easily from the
corresponding heavy-traffic limits in
\cite{CoffmanPuhalskiiReiman1991, CoffmanReiman1983}, and the
ergodicity of $(R(t))$.

Interestingly, it was discovered in the experiments that a normal
limit law also appears to hold for the total number $F$ of fragments
as $M \rightarrow \infty$ (see for example Figure
\ref{normal-fits-num-frags}). This is not surprising,
since $F$ is the sum over all requests in system of the numbers
$F_i,~1 \leq i \leq R,$ of fragments allocated to requests. The
requests have a mutual dependence, but one whose effect can be
expected to weaken for large $M$. 
As illustrated in
Figure~\ref{gapsearch1}, the mean of
$F_i$ is proportional to $M$. Our experiment indicate that the standard deviation of $F_i$ is also approximately proportional to $M$. Hence, if
$\stackrel{p}{\rightarrow}$ denotes convergence in probability as $M
\rightarrow \infty$, then
\begin{equation}
\label{normal-limit} {F - \beta M^2 \over \theta M^{3/2}}
\stackrel{p}{\rightarrow} {\mathcal{N}}(0,1)
\end{equation}
with $\beta \approx 1.5$ and $\theta \approx 0.9$ 
formalizes the normal-limit law suggested by our
experiments.

Note the two departures from the standard
Central Limit Theorem set-up. First, consistent with the linear
fragmentation observation, individual request fragmentation scales linearly
in $M$, so that the total number of fragments scales as $M^2$.
Restating the limit law in terms of the sums of random variables
($F_i/M$) clearly eliminates this discrepancy.   Second, the
number of channels ($R$) is random and satisfies the normal limit law discussed above.  Thus, the plausibility of
(\ref{normal-limit}) requires an appeal to Central Limit Theorems
for random sums (e.g., see \cite{Feller1966}, p. 258, for an
appropriate version). Finally, Figure \ref{normal-fits-num-frags}
gives some idea of the convergence of the fits to the normal
density; as can be seen, fits for $\alpha \leq 0.1$ are indeed
close to the simulation data.


\begin{figure}[!t]
\begin{center}
\scalebox{0.4}{\includegraphics{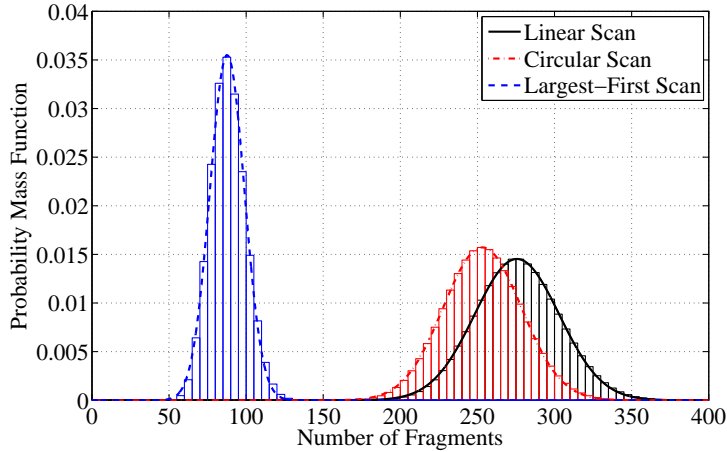}}
 \caption{Distributions of the total number of fragments for $\alpha = 0.1$ and the corresponding Normal fits.}
 \label{fragnormal-CS-LFS}
\end{center}
\end{figure}

We note that the normal  approximations shown for the total number of fragments under LS were also found to hold under CS and LFS.  This is illustrated in Figure~ \ref{fragnormal-CS-LFS}.
\begin{figure}[!t]
\begin{center}
\scalebox{0.4}{\includegraphics{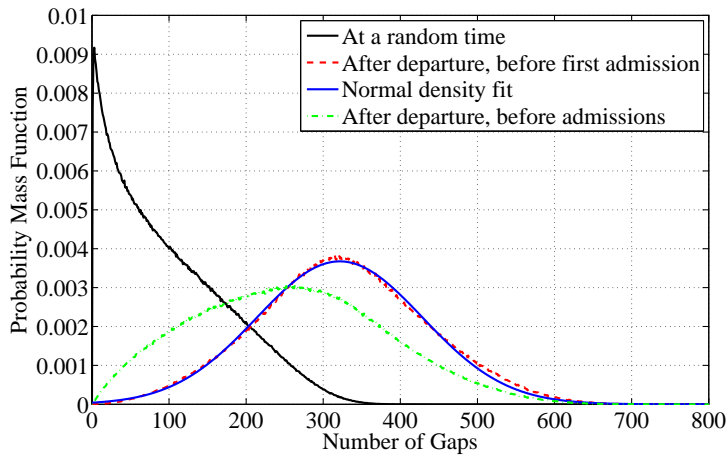}}
 \caption{Distributions of the number of gaps at different epochs for $\alpha = 0.01$.}
 \label{dists-gap}
\end{center}
\end{figure}

The distributions of the number of gaps at different epochs are shown
in Figure \ref{dists-gap}.  The distribution at a random time shows
a decreasing pmf. What appeared to be
yet another normal approximation was discovered when looking at the
{\em first-admission} pmf in Figure \ref{dists-gap};  this is the
distribution as seen by the first admission immediately after a
departure at just those epochs when there is at least one admission.
The third curve is the pmf of the number, $G^-,$ of gaps as seen
right after departures, but before determining whether or not the
head of the queue fits in the total available bandwidth. The fit to
the Normal density is also illustrated in the figure. The
proof of a similar limit law for Renyi's space-filling problem can
be found in \cite{DRlongago}, but extension of known techniques once
again faces the difficult challenges posed by our more difficult
fragmentation problem.
\section{Conclusions}\label{sec:conc}
The results of this paper prepare the ground for further research on
several fronts.  Before listing a number of the more important ones,
we review what we have learned.  Our experiments brought out first
an unexpected reappearance of a {\em 50\% rule} relating the
expected numbers of gaps and channels in the limit of small request
sizes relative to the spectrum size.  In our case, we were able to
prove the limit law.   The next result described a linear
relationship between the maximum request size, $\alpha$, and the
expected number of fragments into which a request was divided at the
time  of allocation. Interestingly, the smaller $\alpha$ was taken,
the greater was the resulting fragmentation of requests.

Our stability
results  established the beginning of a mathematical
foundation of fragmentation processes. Particularly, we showed
that for $\alpha > 1/2$, the fragmentation process is Harris
recurrent. For general $\alpha$, we proved (with considerable
effort) that the total number of fragments is bounded in expected
value.  We examined alternative algorithms for sequencing through the available gaps and showed that using the LFS algorithm leads to significantly less fragmentation than using the LS or CS algorithms. Finally, we exhibited experimentally a limiting,
small-$\alpha$ behavior in which, with appropriate scaling,
distributions tend to Normal.

A broad direction for further research extends the parameters of our
mathematical model.  For instance, while Uniform distributions are
generally the assumption of choice in fragmentation models, it would
be interesting to see what new effects are created by other
distributions of request size, e.g., by varying $a$ in the
generalized uniform distributions on $[0,\alpha]$, with densities
$x^a/\alpha^{a + 1}.$  The exponential residence-time assumption is
likely to yield simplifications to analysis,  but changes
in behavior resulting from other distributions are worth
investigating. Moreover, instead of a system operating at capacity, in which there
is always a request waiting, one could
adopt an underlying, fully stochastic model of demand, e.g., a
Poisson arrival process, as found in \cite{KipnisR1990}.

More realistic, but in all likelihood significantly more difficult
models, would relax the independence assumptions.  A prime example
appropriate for Dynamic Spectrum Access applications would be allowing
residence times to depend on fragmentation, the greater the
fragmentation of a request, the longer its residence time.

The results regarding the performance of the different algorithms imply that the algorithms' design should also be considered carefully. Some examples of algorithms that come to mind will aim to better fit the fragments into the available gaps. A more challenging objective would be to develop algorithms that take into account spectrum sensing capabilities during the gap allocation process.

Finally, another broad and very important avenue of research that
introduces more realistic models discretizes request sizes and the
bandwidth allocation process (as is being done while allocating OFDM subcarriers).  As in other models of fragmentation,
the continuous limit represented in this paper may conceal important
effects, or, conversely, it may introduce effects not present in
discrete models.  We are actively pursuing this avenue of research.

\section{Acknowledgments}
We would like to thank Charles Bordenave for helpful discussions in relation with Theorem \ref{thm:hajek}.
This work was partially supported by NSF grant CNS-0916263 and CIAN NSF ERC under grant EEC-0812072.

\normalsize

\clearpage

\appendix

\section{Tool validation}
\label{app:validation}
In the first test of the simulation tool, we allowed arrivals, discretized
the spectrum size, and fixed the requested bandwidth, in such a way that
the system reduced to a conventional $M/M/k$ queue.  We then
matched simulation results with those obtained from classical formulas.
In all cases checked, the error compared to the exact results was negligible.

Keeping with our continuous model operating at capacity
with random bandwidth requests, one finds that
there are very few explicit results for measures of interest, even
for the process $R(t)$ for the number of channels in the system. Those that do
exist can be found in \cite{KipnisR1990} along with the elegant
formula below, which unfortunately requires calculations that are
rarely tractable.

\begin{theorem}[Kipnis and Robert \cite{KipnisR1990}]
Let $S_n$ denote the sum of $n$
i.i.d.\ request sizes. Then, the maximum throughput (expected number of requests in the system) is

\[
{\mathbb{E}}(R)  =  {\frac{1}{ \sum_{n=1}^{+ \infty} (1/n) \P (S_n \leq 1 <
S_{n+1})} } =  {\frac{1}{\sum_{n=2}^{+ \infty} (1/n(n-1)) \P (S_n > 1)} }
\]

\end{theorem}

A system with requests that are uniformly distributed on $[0,1]$ is one case that admits of a simple
formula. Here, the density of $S_n$ on $[0,1]$ is $z^{n-1}/(n-1)!$,
and so an easy calculation gives
$$ {\mathbb{E}}(R) = \frac{1}{ e - 2} = 1.392\ldots $$
Transform methods are used in \cite{KipnisR1990} to evaluate ${\mathbb{E}}(R)$
for other values of $\alpha$. In particular, computations are based
on the (numerical) inversion of a Fourier-Laplace transform. To evaluate the
accuracy of our simulator, we checked our experimental results
against the computations  for the values of $\alpha$ shown in Table
\ref{kr_comparison}.  As can be seen, the numerics agree well with
simulations.


\begin{table}[!t]
\renewcommand{\arraystretch}{1.3}
\caption{Averages $\overline{R}$, computed from our
experiments, vs. the corresponding values ${\mathbb{E}}(R)$,
obtained from the numerical computations in \cite{KipnisR1990}, for various values of $\alpha$.
Except for the case $\alpha = 1$, the latter results were given to
two decimal digits only.} \label{kr_comparison} \centering
\begin{tabular}{|c|c|c|c|c|c|c|}
\cline{1-3} \cline {5-7}
\bfseries $\alpha$ & ${\mathbb{E}}(R)$ & $\overline{R}$
& &\bfseries $\alpha$ & ${\mathbb{E}}(R)$ & $\overline{R}$\\
\cline{1-3} \cline{5-7} 0.05    &   39.51   &   39.325  &   &   0.55
&   2.9   & 2.892   \\  \cline{1-3} \cline{5-7} 0.1     &   19.4
&   19.317  &   &   0.6     &   2.54  & 2.541   \\  \cline{1-3}
\cline{5-7} 0.15    &   12.69   &   12.653  &   &   0.65    &   2.26
& 2.261   \\  \cline{1-3} \cline{5-7} 0.2     &   9.34    &   9.306
&   &   0.7     &   2.04  & 2.039   \\  \cline{1-3} \cline{5-7} 0.25
&   7.32    &   7.303   &   &   0.75    &   1.87  & 1.868   \\
\cline{1-3} \cline{5-7} 0.3     &   5.98    &   5.961   &   &   0.8
&   1.73  & 1.731   \\  \cline{1-3} \cline{5-7} 0.35    &   5.01
&   5.003   &   &   0.85    &   1.62  & 1.621   \\  \cline{1-3}
\cline{5-7} 0.4     &   4.28    &   4.275   &   &   0.9     &   1.53
& 1.531   \\  \cline{1-3} \cline{5-7} 0.45    &   3.71    &   3.709
&   &   0.95    &   1.45  & 1.456   \\  \cline{1-3} \cline{5-7} 0.5
&   3.29    &   3.281   &   &   1       &   1.392  & 1.392   \\
\cline{1-3} \cline{5-7}
\end{tabular}
\end{table}

\section{Proof of Theorem 4}
\label{app:proof}
%

\noindent
For any $0 < \varepsilon \leq \lambda$,
\begin{multline} \label{eq:0}
    \E_y \left( e^{\varepsilon f(Y_{n+1})} \right)
    = \E_y \left( \E_{Y_n} \left( e^{\varepsilon (f(Y_{1}) - f(Y_0))}
    \right) e^{\varepsilon f(Y_n)} \ind{f(Y_n) \leq K} \right)\\
    + \E_y \left( \E_{Y_n} \left( e^{\varepsilon (f(Y_1) - f(Y_0))}
    \right) e^{\varepsilon f(Y_n)} \ind{f(Y_n) > K} \right).
\end{multline}
Since by assumption, $f(Y_{1}) - f(Y_0)$ under $\P_{y}$ is
stochastically dominated by $Z$ for every $y$, one gets
\[
    \E_{Y_n} \left( e^{\varepsilon (f(Y_{1}) - f(Y_0))} \right) \leq \E \left( e^{\varepsilon Z} \right) = \E \left( e^{\lambda Z} \right) \stackrel{\text{def.}}{=} D
\]
and therefore
\begin{equation} \label{eq:1}
    \E_y \left( \E_{Y_n} \left( e^{\varepsilon (f(Y_{1}) - f(Y_0))} \right) e^{\varepsilon f(Y_n)} \ind{f(Y_n) \leq K} \right) \leq D e^{\varepsilon K}.
\end{equation}
For the second term, we apply Lemma~\ref{lemma} to the random
variable $f(Y_1) - f(Y_0)$ under $\P_{Y_n}$:
\[
    \E_{Y_n} \left( e^{\varepsilon (f(Y_1) - f(Y_0))} \right) \leq 1 + \varepsilon \E_{Y_n} (f(Y_1) - f(Y_0)) + c \varepsilon^2.
\]
Thus, on the event $\{ f(Y_n) > K \}$, one gets
\begin{equation} \label{eq:rho}
    \E_{Y_n} \left( e^{\varepsilon (f(Y_1) - f(Y_0))} \right)
    \leq 1 -\gamma \varepsilon + c \varepsilon^2 \stackrel{\text{def}}{=} \rho_\varepsilon
\end{equation}
and finally
\begin{equation} \label{eq:2}
    \E_y \left( \E_{Y_n} \left( e^{\varepsilon (f(Y_{1}) - f(Y_0))} \right) e^{\varepsilon f(Y_n)} \ind{f(Y_n) > K} \right)
    \leq \rho_\varepsilon \E_y \left( e^{\varepsilon f(Y_n)} \right).
\end{equation}
Gathering the two bounds~\eqref{eq:1} and~\eqref{eq:2}
in~\eqref{eq:0} finally gives
\[ \E_y \left( e^{\varepsilon f(Y_{n+1})} \right)
\leq \rho_\varepsilon \E_y \left( e^{\varepsilon f(Y_n)} \right) + D
e^{\varepsilon K} \] which leads by induction to
\[ \E_y \left( e^{\varepsilon f(Y_{n})} \right) \leq \rho_\varepsilon^n e^{\varepsilon f(y)} + \frac{1-\rho_\varepsilon^{n+1}}{1-\rho_\varepsilon}D e^{\varepsilon K}. \]
Let $\eta = \big (\varepsilon / (2\gamma) \big) \wedge \lambda$.
From the definition of $\rho_\varepsilon$ in ~\eqref{eq:rho}, one
can easily check that $0 < \rho_\eta < 1$. Therefore, choosing
$\varepsilon = \eta$ gives the result.

\end{document}